# Artificial Neural Network and Its Application Research Progress in Chemical Process

Li Sun, Fei Liang, Wutai Cui

School of Chemical Engineering, East China University of Science and Technology, Shanghai 200237, China

**Abstract**  Most chemical processes, such as distillation, absorption, extraction, and catalytic reactions, are extremely complex processes that are affected by multiple factors. The relationships between their input variables and output variables are non-linear, and it is difficult to optimize or control them using traditional methods. Artificial neural network (ANN) is a systematic structure composed of multiple neuron models. Its main function is to simulate multiple basic functions of the nervous system of living organisms. ANN can achieve nonlinear control without relying on mathematical models, and is especially suitable for more complex control objects. This article will introduce the basic principles and development history of artificial neural networks, and review its application research progress in chemical process control, fault diagnosis, and process optimization.

## 1 Introduction

  The production of any product in a chemical plant has to undergo a series of chemical processes such as reaction and separation. Among them, there are many batch or semi-continuous processes with different dangerous conditions such as high temperature and high pressure [1-3]. As the scale of chemical production equipment continues to expand and the degree of automation continues to increase, people have higher and higher requirements for the accuracy and sensitivity of the performance control of industrial processes. However, the traditional control technology has been difficult to meet the requirements. As an important branch of artificial intelligence, artificial neural network has been more and more widely used in chemical production with its intelligent characteristics such as self-adaptation and self-learning, providing practical solutions for the precise and rapid control of complex production equipment. Specifically, artificial neural networks are mainly used in the following aspects of chemical process: fault diagnosis, control and optimization of process parameters, product quality control, physical property estimation, etc. [4-6]. This article will introduce the principle and development history of artificial neural network, and summarize its application research progress in chemical process control, fault diagnosis and process optimization.

**2 Principles and development history of artificial neural networks**

The artificial neural network imitates the human brain neuron network and abstracts it, and then establishes a certain mathematical model, and processes information by adjusting the interconnection between a large number of nodes in the model [7,8]. It has self-adaptive and self-learning functions, and is especially suitable for complex nonlinear information processing systems. When the input value is $x_i$ (i=1,2,•••,n), the output is y, and the relationship between y and $x_i$ is:

$$y = f(s_j) \tag{1}$$

$$s_j = \sum_{i=1}^{n} w_{ji} x_i - \theta_j \tag{2}$$

where $\theta_j$ is the threshold, and $w_{ji}$ is the connection coefficient. The output function f has many forms, the common ones are: proportional function, quadratic function, hyperbolic function, m-type function, Y-type function, etc.

Each node in the neural network has a state variable. The node i and the node j are connected by the connection weight coefficient $w_{ji}$, and each node has a threshold $\theta_j$ and a nonlinear transformation function $f$.

The neuron biology model (M-P model) proposed by McCulloch and Pitts [9] and the Hebb rule proposed by Hebb [10] can be regarded as the beginning of artificial neural networks. The groundbreaking papers of Hopfield [11-13], Kauffman [14], LeCun [15,16] and Hinton [17-19] are the theoretical basis for artificial neural networks to mature. They are ubiquitous in the modeling of neural and gene networks, and they are also indispensable tools in computer science. Hodgkin and Huxley [20] established the famous nonlinear dynamic differential equation, namely the H-H equation. This equation can be used to describe the nonlinear phenomena that occur in the nerve membrane, such as self-excited oscillation, chaos, and multiple stability problems. Rosenblatt [21] proposed the Perceptron model, which is the first physically constructed artificial neural network with learning ability; Widrow [22] proposed the adaptive (Adaline) linear component model, a continuous self-adapt to the linear cell neural network model. In short, through the joint efforts of many scientists, artificial neural networks have entered the stage of commercial application.

The artificial neural network can achieve the approximation of any nonlinear mapping through learning, and it can be applied to the identification and modeling of nonlinear systems without being restricted by the nonlinear model [23]. Its fault tolerance is reflected in the loss of a part of the system. , It will not affect the overall activities [24]. Artificial neural network-based artificial intelligence has familiarity recognition capabilities, classification capabilities, error correction capabilities, and time series retention capabilities. Therefore, it has been widely used in various complex scenarios. The main application areas include economic forecasting [25-27], signal processing [28-30], disease diagnosis [31-33], intelligent driving [34-36], process control and optimization [37-39], image processing [40-42], etc.

According to their different functions, artificial neural networks can be divided into feedforward networks and feedback networks [43]. The main purpose of the feedforward neural network is to learn and recognize. It has a strong recognition ability and can recognize complex molecular states and detect and recognize molecular structures in complex environments. The main function of the feedback neural network is to recurse the network information in stages. After the initial information is input, the information state can be transferred layer by layer, so that the entire network state can reach a dynamic balance. At the same time, through the real-time feedback of the feedback network, the information can be transmitted to various areas, and the content of the information can be output in the form of data, and finally integrated output through the output terminal [44]. Convolutional neural network is another extremely powerful network. It was first proposed by LeCun [45] as a classifier for image recognition. Its basic structure mainly includes input layer, convolutional layer, pooling layer, fully connected layer and output layer. GoodFellow [46] proposed a neural network-based generative model (GAN) in 2014, which inverted the structure of traditional neural networks: its input is a set of low-dimensional noise, and the output is a synthetic image that can be faked. On this basis, Gao et al. proposed an improved GAN model [47], which uses Wasserstein distance to replace the traditional KL distance, and uses a game theory model for training, which improves the stability and convergence of GAN.

## 3 Application research progress of artificial neural network in chemical industry

Artificial neural network has been successfully applied to many fields of chemical process. This article will review its application research progress in chemical process control, fault diagnosis, process optimization and product quality control.

**3.1 Application of neural network in chemical process control**

The heat exchanger is an important chemical equipment used to control the temperature of the material in the rectification system and the reaction system. Due to the nonlinear behavior and complexity of the heat exchanger control process, the traditional PID control method is being replaced by the predictive control based on the neural network model [48]. Vasičkaninová et al. [49] used a neural network predictive control (NNPC) structure to control the heat transfer process. A neural network model is used to predict the future behavior of a controlled process with distributed parameters. In order to test its control efficiency, the tube heat exchanger is selected as the controlled object. The control goal is to maintain the temperature of the heated outlet stream at the required value and minimize energy consumption. The results show that the NNPC of the heat exchanger consumes less heating medium compared with the classic PID (Proportional-Integral-Derivative) control, which proves the effectiveness and superiority of NNPC. Longo et al. [50] proposed an artificial neural network (ANN) model to predict the boiling heat transfer coefficient of refrigerant in a brazed plate heat exchanger (BPHE). The model takes into account the influence of plate geometry, operating conditions and refrigerant characteristics, and the average error (MAPE) of the predicted value is 4.8%. Compared with most of the most advanced analysis and calculation programs available in the public literature for internal boiling of BPHE, the ANN model shows better predictive ability.

Compared with traditional PID control, Model Predictive Control (MPC) has higher operational efficiency. Shin et al. [51] used an artificial neural network (ANN) model to replace the existing linearization model, and used Aspen HYSYS to simulate the de-propanizer. They consider all feasible operating scenarios to generate a large amount of dynamic simulation data and use it for the training and testing of artificial neural networks. The results show that MPC requires shorter settling time and rise time than PID control, which proves that it has more sensitive control performance. Osuolale et al. [52] proposed a strategy based on neural networks and the second law of thermodynamics to model and optimize the energy efficiency of distillation towers. Bootstrap aggregated neural network can enhance the accuracy and reliability of the model. Aspen HYSYS is used for the distillation simulation of the methanol-water and benzene-toluene separation binary system, which ultimately reduces utility consumption by 8.2% and 28.2%, respectively. Yang et al.

[53] used a 3-layer BP network model to simulate and predict the process of ultrasonic-enhanced supercritical extraction of flavonoids from toona sinensis leaves. The effects of various factors such as extraction temperature, extraction pressure, fluid flow, amount of entrainer, extraction time and ultrasonic power on the extraction rate of total flavonoids are discussed separately, and the errors of the results obtained are small. Hu et al. [54] used artificial neural networks to simulate the start-up process of catalytic distillation towers. The learning algorithm of the artificial neural network used in this study is the L-M method. The input parameters are the composition and total moles of the initial feed liquid in the tower, and the output parameter is the time it takes for the catalytic distillation to start. The prediction performance of the artificial neural network increases with the increase of training data, and the prediction error gradually decreases.

**3.2 Application of neural network in fault diagnosis**

Efficient fault prevention, fault detection and fault diagnosis of processes and devices are necessary conditions for safe, stable and efficient production. Artificial neural network models rely on powerful self-learning and nonlinear mapping capabilities, which are widely used in fault diagnosis and prevention of chemical processes and equipment. Currently, BP networks, radial basis (RBF) networks and adaptive networks are commonly used for fault diagnosis [2]. BP network is a kind of global approximation neural network, and the ownership value and threshold value of the network need to be revised during the training process. Downs et al. [4] used BP neural network(BPNN) to simulate a more complex chemical process, which mainly includes 5 typical units: reactor, condenser, gas-liquid separator, desorption tower, and circulating compressor. There are a total of 4 reactions, producing 2 products and one by-product. The 15 known fault numbers are represented by a 4-dimensional vector, and disturbances are added. Take out 500 sets of data to train the neural network, and then take 100 sets as test data. Through Levenberg-Marquardt back-propagation algorithm for training, BPNN shows high fault recognition ability. Gu [56] used BP neural network and RBF neural network to detect and diagnose faults, and found that in the numerical simulation examples, the process relationship is static and linear. BP network has a good effect on fault detection. Manssouri et al. [57] established a reliable model ANN-ELM based on the ELM (Extreme Learning Machine) type of artificial neural network, which can distinguish between normal and abnormal patterns. It is applied to the distillation column with toluene/methylcyclohexane. All relevant inputs are: heating power, preheating power, reflux rate, feed rate, pressure drop and preheating temperature; the output is the tower top temperature. After training and testing in a database of 1000 samples, the results show that the prediction accuracy of the ELM model is very good. When the number of neurons in the hidden layer was 30, a low RMSE value (RMSE = 0.0168) was recorded during the test phase. The ANN-ELM predictive model is most suitable for the normal mode modeling of the variable operating point of the automatic continuous distillation column, and can be used for online detection and diagnosis.

### 3.3 Application of neural network in process optimization

The production process of a chemical product often involves multiple influencing factors. The functional relationship between product yield and purity (output parameters) and these factors (input parameters) is complex and changeable, and cannot be correlated by mathematical models. Neural networks have a typical black-box nature. They can automatically extract "reasonable" solving rules by learning a set of examples containing correct answers, and establish a mapping relationship between output and input. It has been applied to the process optimization of chemical products. Kang et al. [58] optimized the process of polyvinylidene fluoride/polypropylene gradient composite filter media based on BP neural network. Taking the fiber membrane filtration resistance as the target value, MATLAB is used to construct a feed-forward neural network, and the algorithm can master the calculation ability through learning, training, and testing of sample data. Take voltage, receiving distance and injection speed as three input units, set up four hidden units, and take filtering resistance as output unit. Use the newff function to create a network object for pre-feedback training. Set the network parameters, and feedback the result every 400 times of training. The Sigmoid excitation function is used to calculate the output value of each layer. Select 70% of the data for training, 15% of the data for correction, and 15% for testing to get the fitting curve. The optimized process parameters are the voltage of 30kV, the receiving distance of 16.8cm, and the flow rate of 1.6mL/h. The relative error of the BP neural network is 1.99%, and the resistance prediction value is 81.25 Pa, which has high prediction accuracy. Zhu et al. [59] used a new adaptive genetic neural network to optimize the process of extracting polysaccharides from burdock solid fermented Ganoderma lucidum. MATLAB R2016B was used to construct an adaptive genetic neural network, and the test data extracted from Ganoderma lucidum polysaccharide were used to train the neural network. Setting the initial interval of the ownership value threshold [-10,10] and the variable accuracy of $10^{-4}$, the adaptive genetic neural network algorithm was used to fit the experimental value of the polysaccharide content. The results show that the 13 sets of predicted values of the genetic neural network model are almost completely fitted with the experimental ones, and the adaptive genetic neural network algorithm has higher prediction and optimization capabilities than the regression analysis method.

Su et al. [60] used a radial basis neural network model to optimize the injection molding process of the light guide. Taking the light guide strip of automobile front combination lamp as an example, the optimal Latin hypercube sampling method is selected to obtain the sample. Five parameters (melt temperature, mold temperature, holding time, holding pressure and cooling time) were selected as the input layer and two parameters (minimum volume shrinkage rate and minimum sink mark index) as the output layer, and a radial basis basis (RBF) neural network model was constructed. The Insight optimization module is used to obtain a set of parameters for optimal injection molding process, and the actual simulation results are basically consistent with the predicted ones, which effectively improves the molding quality. Zhang et al. [61] used artificial neural network (GA-ANN) to optimize the extraction process of Tibetan tea polysaccharide (TTP) and evaluated its in vitro antioxidant activity. Taking liquid-to-material ratio, extraction temperature, and extraction time as input parameters, and TTP extraction rate as output parameters, the response surface method (RSM) and genetic algorithm-artificial neural network (GA-ANN) were used to optimize the extraction process. Du et al. [62] applied artificial neural network model and optimized genetic algorithm to optimize and predict the most suitable protoplast preparation process of Paecilomyces tenuipes. The protoplasts of Paecilomyces tenuipes were prepared according to the optimal preparation process obtained by the above optimization. The average production of protoplasts in 5 parallel experiments was $4.4 \times 10^7$ cells/mL, and the error with the predicted value of the ANN model was 0.23%.

## 4 Conclusion

The ANN has a self-learning function and a typical black-box nature. It can automatically extract "reasonable" solving rules by learning a set of examples containing correct answers, and establish a mapping relationship between output and input. ANN has been successfully applied to various segments of the chemical process. This article briefly describes the principle and development process of ANNs, and summarizes its application research progress in three chemical sub-fields.

With the further development of computer technology and the continuous innovation of neural network algorithms, the application of ANNs in the chemical industry will become more extensive.